\documentclass[%
reprint,
superscriptaddress,
amsmath,
amssymb,
aps,
]{revtex4-1}
\usepackage{graphicx}
\usepackage{dcolumn}
\usepackage{bm}
\usepackage[mathlines]{lineno}
\usepackage{hyperref}
\usepackage{xcolor}
\hypersetup{
  colorlinks=true,
  linkcolor=blue,
  citecolor=blue,
  urlcolor=blue,
}
\usepackage{color,soul}
\usepackage[T1]{fontenc}
\usepackage{siunitx}
\DeclareSIUnit\torr{Torr}
\DeclareSIUnit\oersted{Oe}

\begin{document}

\preprint{APS/123-QED}

\title{Controllable frequency tunability and parabolic-like threshold current behavior in spin Hall nano-oscillators}

\author{Arunima TM}
\affiliation{Department of Physics, Indian Institute of Technology Roorkee, Roorkee 247667, India.}
\author{Himanshu Fulara}
\email{himanshu.fulara@ph.iitr.ac.in}
\affiliation{Department of Physics, Indian Institute of Technology Roorkee, Roorkee 247667, India.}

\date{\today}

\begin{abstract}
We investigate the individual impacts of critical magnetodynamical parameters—effective magnetization ($\mu_0 M_{\text{eff}}$) and magnetic damping ($\alpha$)—on the auto-oscillation characteristics of nano-constriction-based Spin Hall Nano-Oscillators (SHNOs). Our micromagnetic simulations unveil a distinctive non-monotonic relationship between current and auto-oscillation frequency in out-of-plane magnetic fields. The influence of effective magnetization on frequency tunability varies with out-of-plane field strengths. At large out-of-plane fields, the frequency tunability is predominantly governed by effective magnetization, achieving a current tunability of 1 GHz/mA—four times larger than that observed at the lowest effective magnetization. Conversely, at low out-of-plane fields, although a remarkably high-frequency tunability of 4 GHz/mA is observed, the effective magnetization alters the onset of the transition from a linear-like mode to a spin-wave bullet mode. Magnetic damping primarily affects the threshold current with negligible impact on auto-oscillation frequency tunability. The threshold current scales linearly with increased magnetic damping at a constant out-of-plane field but exhibits a parabolic behavior with variations in out-of-plane fields. This behavior is attributed to the qualitatively distinct evolution of the auto-oscillation mode across different out-of-plane field values. Our study not only extends the versatility of SHNOs for oscillator-based neuromorphic computing with controllable frequency tunability but also unveils the intricate auto-oscillation dynamics in out-of-plane fields.
\end{abstract}

\pacs{Valid PACS appear here}
\maketitle  
\section{\label{sec:level1}Introduction}

Spin Hall nano-oscillators (SHNOs)~\cite{demidov2012ntm,liu2012prl,duan2014ntc,demidov2014apl,Chen2016procieee,Awad2016NatPhys,ren2023natcom,fulara2020natcomm,montoya2023easy} are prominent spintronic devices that leverage the spin Hall effect (SHE) phenomenon ~\cite{hirsch1999prl,hoffmann2013ieeem,sinova2015rmp} within the heavy metal layer to convert a lateral direct charge current into a transverse pure spin current. This spin current then drives magnetization auto-oscillations in nanoscopic regions within the adjacent ferromagnetic layer through the transfer of spin angular momentum~\cite{demidov2012ntm,demidov2014apl,Awad2016NatPhys,Zahedinejad2018APL}. Recently, SHNOs have attracted significant attention due to their advantages in nanofabrication, rapid and ultra-broadband microwave frequency tunability, compatibility with complementary metal-oxide-semiconductor (CMOS) technology, and emerging applications in neuromorphic computing and magnonics~\cite{demidov2014apl,Mazraati2016apl,Chen2016procieee,divinskiy2017apl,Tarequzzaman2019CommPhys,Zahedinejad2018APL,Fulara2019SciAdv,behera2022energy}. Notably, among various SHNO device geometries, nano-constriction-based SHNOs offer distinct advantages, including flexible device layouts, direct optical access to auto-oscillating regions, and voltage control of the constriction region~\cite{demidov2014apl,Awad2016NatPhys,Zahedinejad2019NatN,Zahedinejad2022natmat,fulara2020natcomm,muralidhar2022optothermal}. Their highly non-linear properties and in-plane current flow facilitate the mutual synchronization of multiple SHNOs in both linear chains and two-dimensional arrays, enabling high-quality microwave signal generation and scaling neuromorphic computing to large oscillator networks ~\cite{Awad2016NatPhys,Zahedinejad2019NatN,Zahedinejad2022natmat,kumar2023robust}. The precise control of auto-oscillation frequency over a wide range plays a pivotal role in controlling the dynamical couplings between oscillators, offering a potential avenue for facilitating synaptic communication among artificial neurons. To achieve learning in oscillator-based neuromorphic computing, robust frequency tunability is crucial, especially at higher operating frequencies ~\cite{torrejon2017neuromorphic,romera2018nature,markovic2020physics}. The agility to control frequency tunability is also beneficial for communication systems where specific frequency bands are allocated for different applications. The degree of frequency tunability depends on magnetodynamical properties, drive current, and the orientation and strength of the applied magnetic field~\cite{slavin2005ieeem,Dvornik2018PRA}. In a recent study, Haidar et al.~\cite{Haidar2021apl} reported the impact of NiFe alloy composition on the auto-oscillation properties of nano-constriction-based Ni$_{\text{100-x}}$Fe$_{\text{x}}$/Pt SHNOs under in-plane fields. The study revealed that Fe-rich devices require higher current densities for driving auto-oscillations at higher frequencies, raising a fundamental question about how damping and saturation magnetization independently influence these auto-oscillation properties. Despite numerous studies on the auto-oscillation behavior of nanoconstriction SHNOs, a comprehensive understanding of the independent roles of key magnetodynamical parameters, such as effective magnetization ($\mu_0 M_{\text{eff}}$) and magnetic damping ($\alpha$) on the auto-oscillation properties remains elusive. 

In this study, we systematically investigate the individual impacts of effective magnetization and magnetic damping on the magnetization auto-oscillations of SHNOs. Our observations reveal qualitatively similar auto-oscillation behavior for different effective magnetization values at a fixed out-of-plane field. However, the frequency tunability controlled by effective magnetization exhibits significant variations at different out-of-plane fields. Conversely, magnetic damping primarily influences the threshold current with minimal impact on frequency tunability. Additionally, the threshold current of auto-oscillation at different out-of-plane fields exhibits an intriguing parabolic-like behavior, indicative of a qualitatively different spatial evolution of the auto-oscillating mode within distinct field energy landscapes.

\section{\label{sec:level2}Micromagnetic Simulations}

\begin{figure}[t]
\centering
\includegraphics[width=8.5cm]{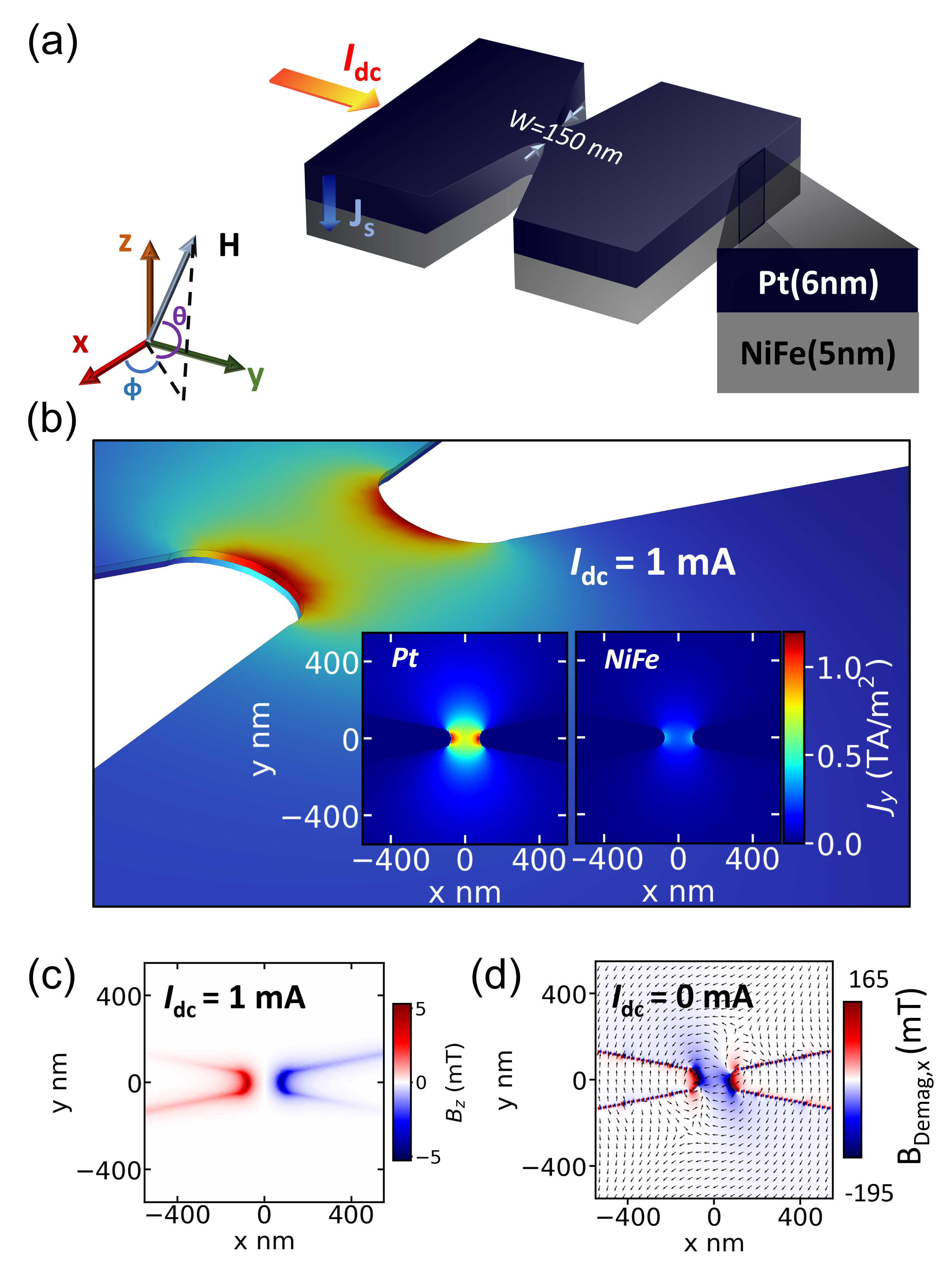}
\caption{Device schematic, current density, Oersted field, and demagnetizing field (a) Schematic of a nano-constriction-based Pt(6 nm)/Pt(5 nm) spin Hall nano-oscillator (SHNO), with a constriction width of $w$ = 150 nm. (b) Distribution of lateral charge current density in the Pt/Py nano-constriction for an input current, $I_{dc}$ = 1 mA. The inset illustrates the simulated current density in the individual Pt and Py layers. (c) Estimation of the out-of-plane component of the Oersted (Oe) field in the Py layer for $I_{dc}$ = 1 mA. (d) Distribution of demagnetizing field in the Py layer for $I_{dc}$ = 0 mA.}
\label{fig:1} 
\end{figure}

Here, we simulate a Pt(6 nm)/Py(5 nm) nano-constriction-based SHNO with an optimized constriction width of 150 nm, as illustrated schematically in Fig. \ref{fig:1}(a). The device geometry is initially modeled using the COMSOL~\cite{COMSOL} Multiphysics Software. Simulations consider the full-scale Pt/Py bilayer with an input current, $I_{dc}$ = 1 mA, and utilize resistivity values of 32.6 $\mu \Omega$-cm for Py and 11.2 $\mu \Omega$-cm for Pt, as reported in experimental study~\cite{demidov2014apl}. Figure \ref{fig:1}(b) shows the charge current density distribution along the y-axis in a 150 nm SHNO obtained from COMSOL simulations. Notably, there is a marked increase in local current density at the constriction region. Due to the disparate electrical resistances of the Pt and Py layers, the current predominantly flows through the Pt metal, as illustrated in the inset of Fig. \ref{fig:1}(b). As a result, in the micromagnetic simulations, we neglect the contribution of the charge current passing through the Py layer to influence the magnetization dynamics. Given that the spin current injected into Py is directly proportional to the current density in Pt, $J_{S}= \Theta_{SH} J_{C}$, the nano-constriction region, characterized by a large current density, defines the active device area, wherein the injected spin current is large enough to excite magnetization dynamics. Figure \ref{fig:1}(c) shows the simulated out-of-plane component of the current-induced Oersted (Oe) field profile in the Py layer for an input current, $I_{dc}$ = 1 mA. Using MUMAX3~\cite{vansteenkiste2014aip}, we then modeled a NiFe layer characterized by an exchange stiffness ($A_{ex}$) of 10 pJ/m and a gyromagnetic ratio ($\gamma$/2$\pi$) of 29.53 GHz/T~\cite{Awad2016NatPhys} within a mesh area measuring 2000 x 2000 nm², divided into a grid size of 512 x 512. The externally applied magnetic field has fixed in-plane ($\phi$) and out-of-plane angles ($\theta$) of 24 and 80 degrees, respectively.

The spatial distribution of demagnetizing fields in the nano-constriction area is illustrated in Fig. \ref{fig:1}(d), obtained through micromagnetic simulations using the MUMAX3 solver~\cite{vansteenkiste2014aip}. The calculations reveal that the nano-constriction edges produce an Oe field and a demagnetizing field opposing the external static magnetic field, leading to a localized decrease in the internal field within the nano-constriction area of the Py film. Our methodology involves running current sweep simulations to investigate the influence of independently varying $M_{eff}$ and $\alpha$ on the magnetization auto-oscillation frequency under both lower and higher applied magnetic fields. Subsequently, Fast Fourier Transform (FFT) calculations were applied to the time-dependent and space-dependent magnetization data to extract frequencies and spatial profiles of the auto-oscillating modes.

\begin{figure}[t]
\centering
\includegraphics[width=8.5cm]{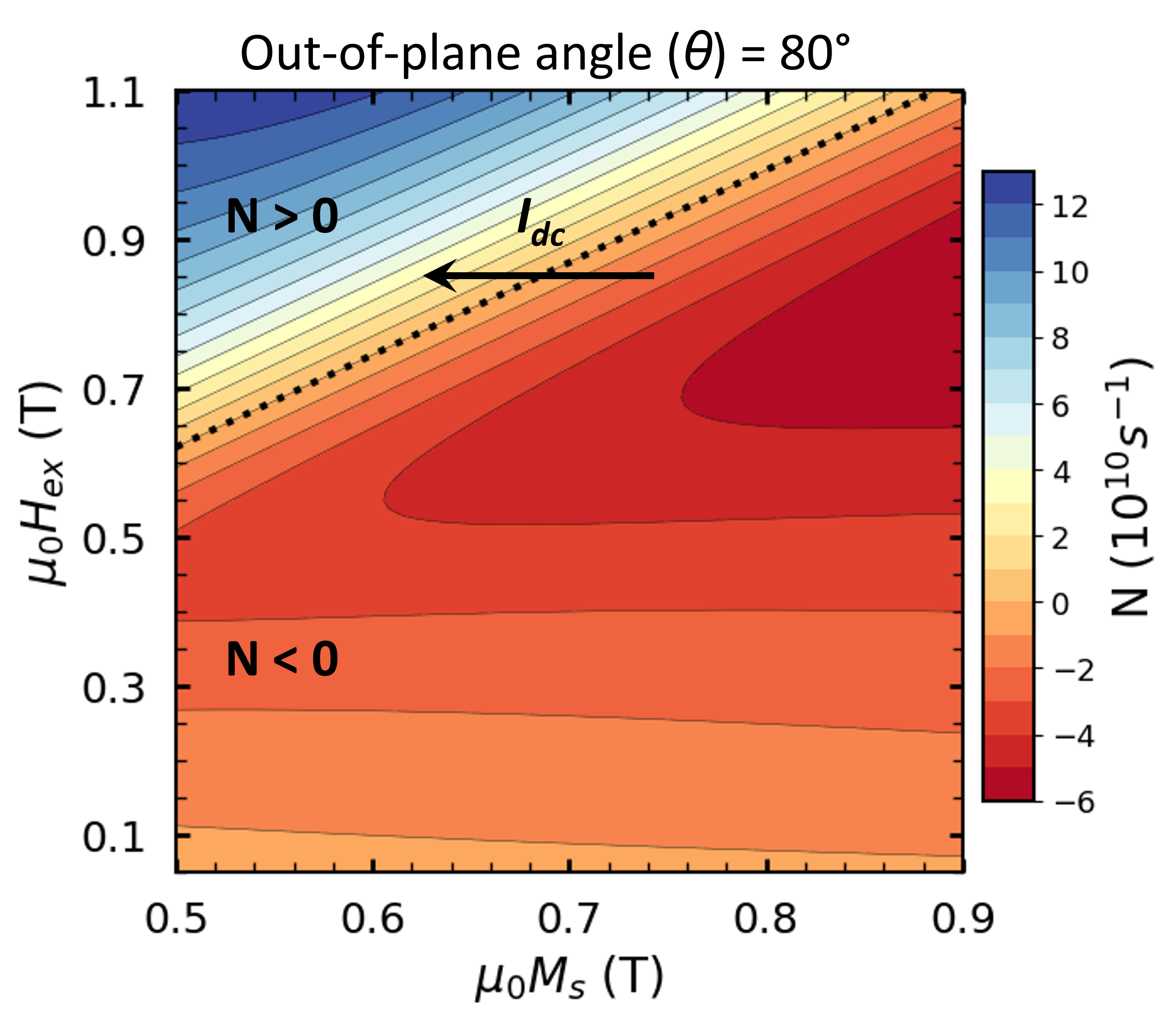}
\caption{Contour plot illustrating the analytically determined non-linearity coefficient ($\mathcal{N}$) for a thin ferromagnetic film, plotted as a function of saturation magnetization and externally applied out-of-plane magnetic field ($\theta =80^{\circ}$). The dotted black line shows the $\mathcal{N}$ = 0 region. The black arrow signifies the enhancement in the precession amplitude with applied current, resulting in an effective reduction of $M_{\text{S}}$ via the projection of magnetization ($M_{\beta}$ = $M_{\text{S}}$ cos $\beta$) along the static field direction. Here, $\beta$ represents the precession angle of magnetization.
}
\label{fig:2}
\end{figure}

\section{Magnetodynamical non-lineraity}

The intricate non-linear magneto-dynamics within patterned magnetic thin films can be analytically characterized by a non-linearity coefficient, denoted as $\mathcal{N}$~\cite{slavin2009}. The magnitude and sign of $\mathcal{N}$ play a pivotal role in determining the strength and nature of magnon-magnon interactions, where positive and negative values correspond to magnon repulsion and attraction, respectively \cite{slavin2009,Dvornik2018PRA}. A negative non-linearity leads to a reduction in the auto-oscillation frequency with increasing amplitude, ultimately causing it to enter the spin-wave band gap. This leads to the self-localization of spin-waves, triggering the excitation of solitonic modes such as spin-wave bullets in in-plane magnetized films \cite{slavin2005prl,mazraati2018pra}, and magnetic droplets in films exhibiting very large perpendicular magnetic anisotropy (PMA) \cite{mohseni2013sc,divinskiy2017prb}. Conversely, large positive non-linearity results in an increase in the auto-oscillation frequency with amplitude, surpassing the ferromagnetic resonance (FMR) frequency and leading to the excitation of propagating spin waves\cite{Slonczewski1999jmmm,madami2011nn,Fulara2019SciAdv}. 

The magnetodynamical non-linearity is primarily governed by material parameters such as saturation magnetization ($M_{\text{S}}$), magnetic anisotropy, and the strength and orientation of applied magnetic fields. To analytically calculate the nonlinear coefficient $\mathcal{N}$ for an in-plane magnetized film, shown in Fig. \ref{fig:2}, we adopted the methodology outlined in Ref. ~\cite{slavin2009,mohseni2018prb}. The results are plotted as a function of $M_{\text{S}}$ and externally applied out-of-plane magnetic fields ($\theta =80^{\circ}$). As shown in Fig. \ref{fig:2}, $\mathcal{N}$ predominantly exhibits negative values (depicted by red regions) at low applied magnetic fields, regardless of variations in $M_{\text{S}}$. However, at higher applied fields, $\mathcal{N}$ undergoes a monotonic transition from negative (red regions) to positive values (illustrated by blue regions), depending on the strength of $M_{\text{S}}$. Notably, it crosses through zero at a specific field strength (indicated by the black dotted line), the precise location of which is influenced by $M_{\text{S}}$. In other words, the reduction in $M_{\text{S}}$ shifts the point of zero non-linearity towards lower values of applied fields. Tailoring non-linearity through the manipulation of effective magnetization and applied magnetic fields in SHNOs should, in principle, offer precise control over frequency tunability and magnetization auto-oscillation dynamics.

\section{Results and Discussion}
\begin{figure*}[t]
\centering
\includegraphics[width=17cm]{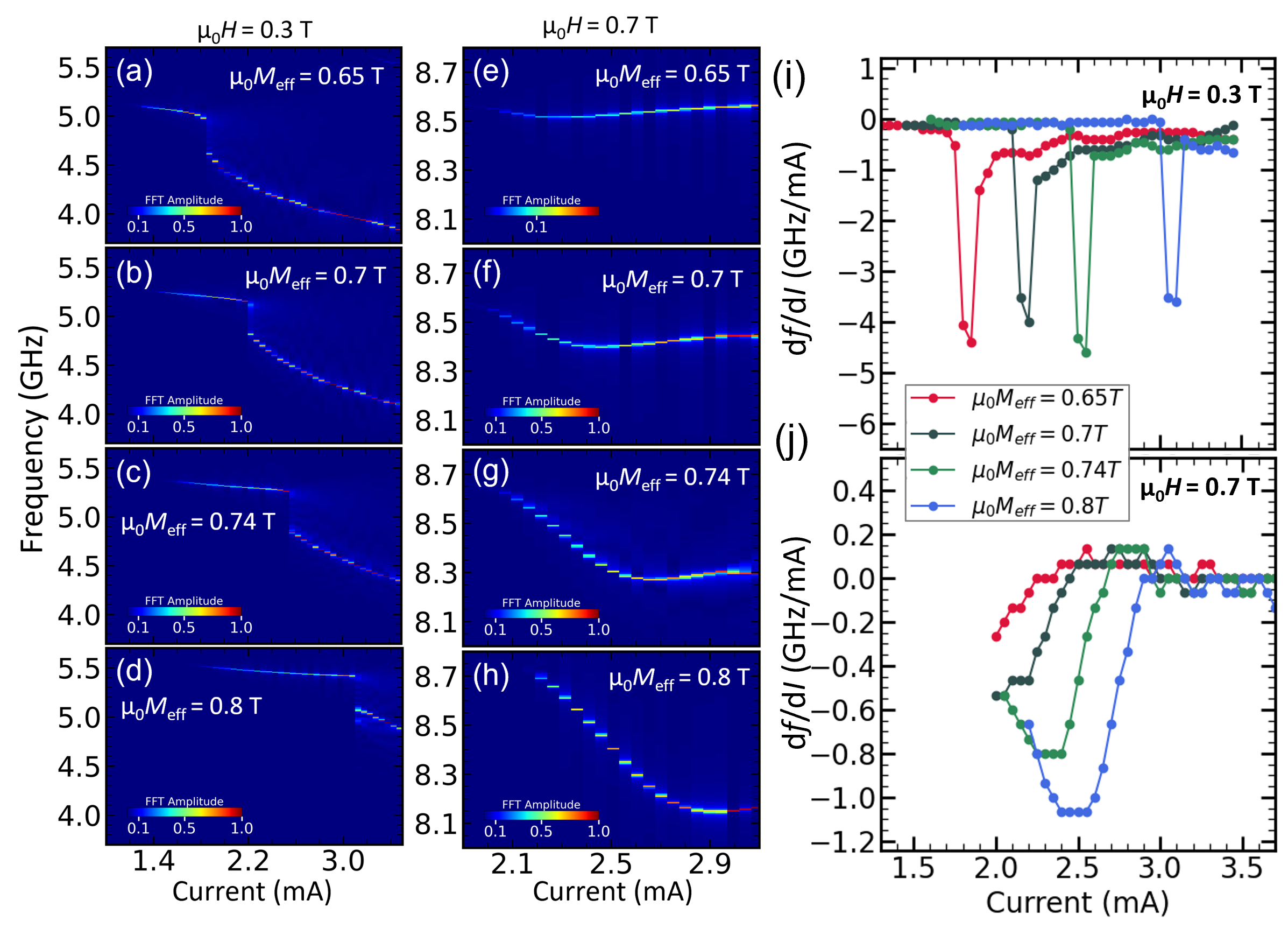}
\caption{Power spectral density obtained from micromagnetically simulated data, illustrating its variation with current for four distinct effective magnetization values under out-of-plane magnetic fields ($\theta$=80$^{\circ}$, $\varphi$=24$^\circ$): (a-d) $\mu_{o}H$ = 0.3 T, (e-h) $\mu_{o}H$ = 0.7 T.  Comparison of auto-oscillation frequency tunability ($df/dI$) as a function of drive current for four distinct effective magnetization values at (i) $\mu_{o}H$ = 0.3 T, (j) $\mu_{o}H$ = 0.7 T
}
\label{fig:3}
\end{figure*}

Figure \ref{fig:3}(a-h) presents micromagnetically simulated current sweep auto-oscillations under two qualitatively different operating regimes for four distinct effective magnetization values, which are practically tunable through the optimization of Fe composition in the Ni$_{\text{100-x}}$Fe$_{\text{x}}$ layer~\cite{Haidar2021apl}. The objective was to discern the influence of effective magnetization on auto-oscillation properties while maintaining other magnetodynamical parameters constant. In Fig. \ref{fig:3}(a-d), at the low applied field of $\mu_{o}H$ = 0.3 T, the auto-oscillations display a typical redshifted frequency vs. current behavior, accompanied by a linear rise in threshold frequency with effective magnetization. A similar red-shifted frequency behavior is experimentally observed in earlier studies on SHNOs~\cite{mazraati2018pra,RajabaliPRA2023}. The frequency of auto-oscillations remains nearly constant at low current values, but at a specific effective magnetization-governed current, there is an abrupt jump in the frequency indicative of the so-called spin-wave bullet—a non-topological self-localized mode that emerges in large negative non-linearity regions~\cite{slavin2009,Dvornik2018PRA,mazraati2018pra}, as shown in Fig. \ref{fig:2}.

In the presence of large out-of-plane field of $\mu_{o}H$ = 0.7 T, as shown in Fig. \ref{fig:3}(e-h), auto-oscillations exhibit a nonmonotonic frequency vs. current behavior, characterized by a red-shifted frequency at lower currents followed by a blue-shifted one at higher current values~\cite{Awad2016NatPhys,Dvornik2018PRA}. This distinct non-monotonic frequency behavior has been extensively reported in recent experimental investigations on SHNOs~\cite{Awad2016NatPhys,AwadAPL2020,Fulara2019SciAdv,RajabaliPRA2023,chen2019pra}. For in-plane magnetized Py thin films, the non-linearity at high applied fields is predominantly negative, and therefore, the auto-oscillation frequency initially experiences a redshift with increasing drive current~\cite{Dvornik2018PRA,AwadAPL2020}. As the drive current increases, the precession amplitude increases, and the demagnetizing field along the static field direction reduces, leading to a situation where the non-linearity becomes zero (refer to Fig. \ref{fig:2}). Consequently, the redshifting of the frequency stops, and the auto-oscillation frequency passes through a minimum where the non-linearity is zero. With any further increase in the drive current, the non-linearity becomes positive, inducing a blue shift in the auto-oscillation frequency~\cite{Dvornik2018PRA,AwadAPL2020,Fulara2019SciAdv}. Similar to the low-field auto-oscillation behavior, this nonmonotonic frequency response at higher fields remains essentially unchanged in nature across all four distinct $\mu_0 M_{\text{eff}}$ values, with only a minor increase in onset auto-oscillation frequency. However, the frequency tunability demonstrates a significant boost at higher effective magnetization values, achieving a fourfold increase in current tunability at $\mu_0 M_{\text{eff}}$ = 0.8 T. As illustrated in Fig. \ref{fig:3}(i), at low fields, effective magnetization primarily influences the onset of the bullet mode by significantly enhancing negative non-linearity while moderately adjusting frequency tunability, reaching a maximum current tunability of 4 GHz/mA. Conversely, at high fields, frequency tunability experiences a fourfold increase with increasing effective magnetization values (refer to Fig. \ref{fig:3}(j)). Our results indicate the pivotal role of effective magnetization in tailoring the magnetodynamic non-linearity at the active dynamical region, thereby influencing both the frequency tunability and the onset of the bullet mode.

\begin{figure}[t]
\centering
\includegraphics[width=8.5 cm]{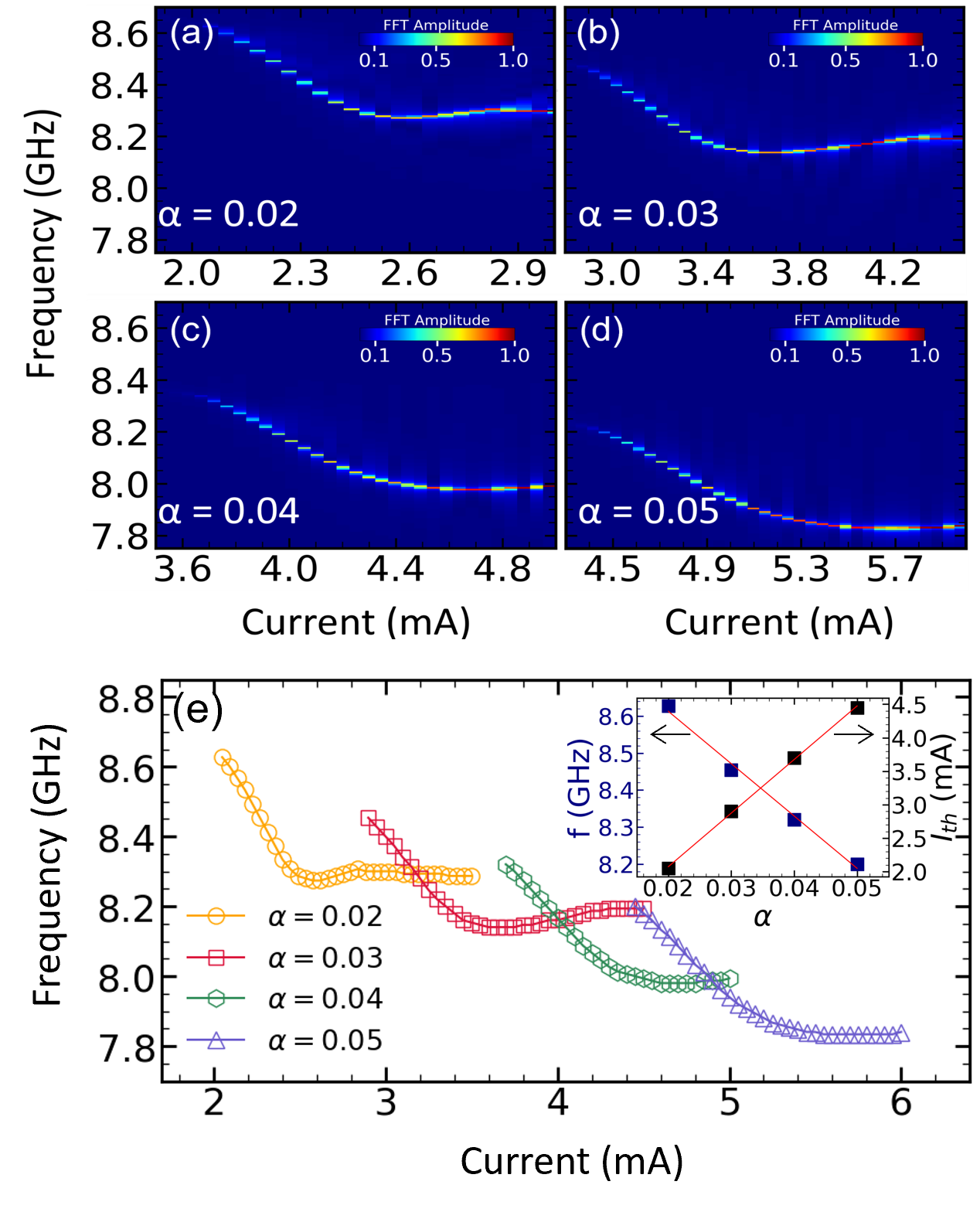}
\caption{Micromagnetically simulated auto-oscillation frequency vs.~drive current for four damping values at a fixed $\mu_0 M_{\text{eff}}$ = 0.74 T : (a) $\alpha$ = 0.02, (b) $\alpha$ = 0.03, (c) $\alpha$ = 0.04, (d) $\alpha$ = 0.05. (e) Comparison of auto-oscillation frequencies as a function of drive current at various damping magnitudes. The inset shows the variation of threshold frequency (blue squares) and threshold current (black squares) as a function of damping, with red lines indicating linear fits.
}
\label{fig:4}
\end{figure}

Subsequently, we explore the critical influence of magnetic damping on the auto-oscillation properties of SHNOs, while maintaining a fixed effective magnetization value of $\mu_0 M_{\text{eff}}$ = 0.74 T. This investigation is confined to high out-of-plane fields, $\mu_{o}H$ = 0.7 T ($\theta$=80$^{\circ}$, $\varphi$=24$^\circ$), where significant variations in effective magnetization-controlled frequency tunability were observed (refer to Fig. \ref{fig:3}(j)). In Fig. \ref{fig:4}(a-d), we present current sweep auto-oscillations at four distinct magnetic damping values ranging from $\alpha$ = 0.02 to 0.05. As shown in  Fig. \ref{fig:4}(e), magnetic damping predominantly influences the threshold current and onset frequency, with negligible impact on auto-oscillation frequency tunability. The qualitatively similar frequency tunability implies that the fundamental nature of auto-oscillations remains unaffected by variation in damping. As illustrated in the inset of Fig. \ref{fig:4}(e), the onset frequency demonstrates a linear decrease with an increase in $\alpha$, while the threshold current increases with increasing $\alpha$ values. Note that we estimated the threshold current by monitoring the magnetization behavior over time. The initial 5 ns were excluded to mitigate transient effects, and the subsequent 15 ns were observed to detect the magnetization auto-oscillations. If the auto-oscillations sustain over time, we identify it as the onset of auto-oscillation. The observed enhancement in threshold current is slightly below the anticipated value of 2.5 times, indicating that the increase in damping is linked with the stronger localization of auto-oscillating mode due to a higher Oe field. The Oe field effectively suppresses the spin-wave well on one side of the nano-constriction while enhancing the localization of the mode on the opposite side~\cite{Dvornik2018PRA}.

\begin{figure}
\centering
\includegraphics[width=8.5 cm]{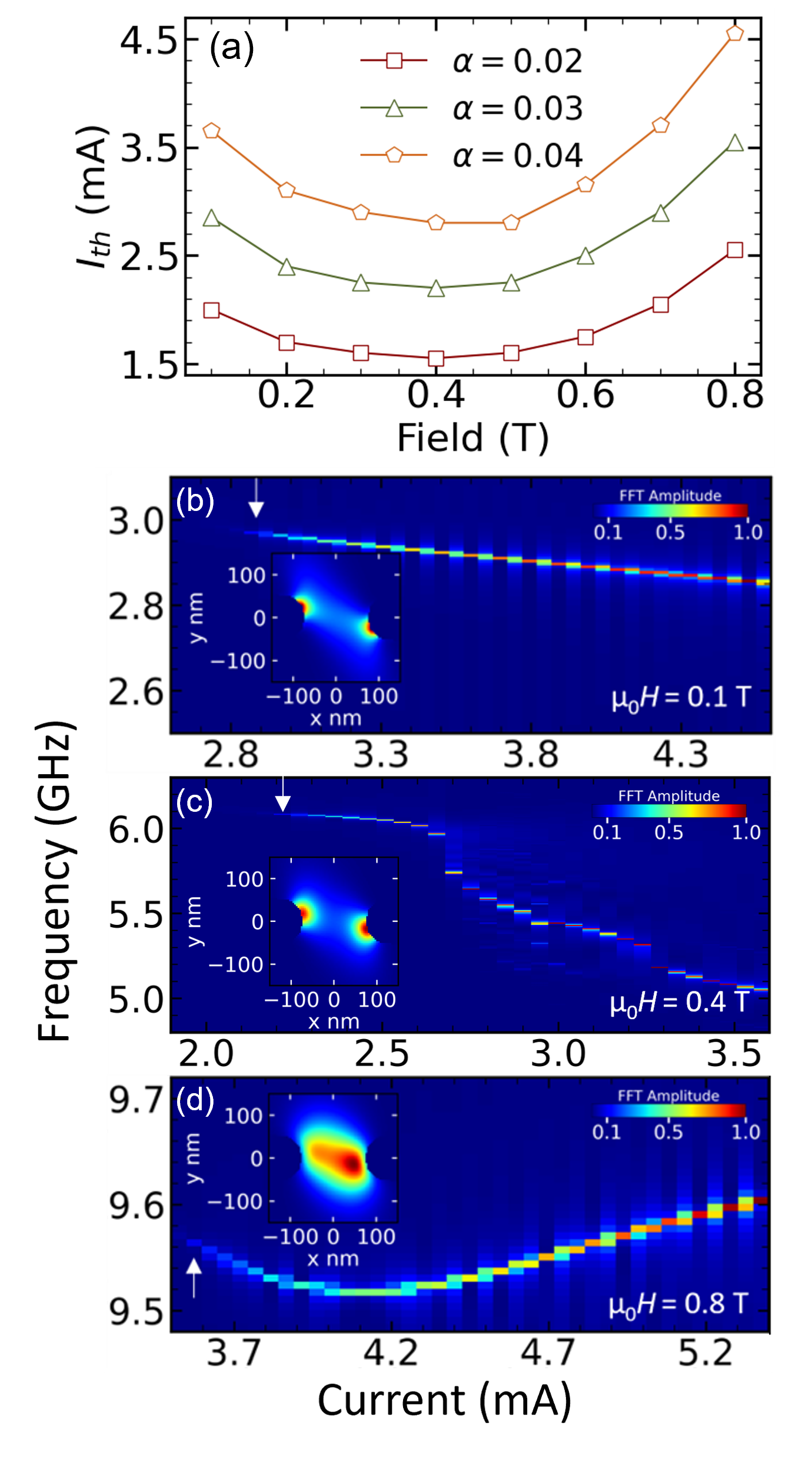}
\caption{(a) Variation of threshold current ($I_{th}$) as a function of applied out-of-plane fields ($\theta$=80$^{\circ}$, $\varphi$=24$^\circ$), calculated for three different damping values. Micromagnetically simulated current sweep auto-oscillations for $\alpha$ = 0.03 subjected to three distinct out-of-plane field strengths: (b) $\mu_{o}H$ = 0.1 T, (c) $\mu_{o}H$ = 0.4 T, and (d) $\mu_{o}H$ = 0.8 T. The inset illustrates the spatial profiles of auto-oscillations simulated at the onset of auto-oscillations for each field strength.
}
\label{fig:5}
\end{figure}

To delve deeper into the impact of magnetic damping across various operational regimes, micromagnetic simulations were conducted, maintaining a constant effective magnetization of $\mu_0 M_{\text{eff}}$ = 0.74 T while varying out-of-plane magnetic fields ($\mu_{0}H$). Figure \ref{fig:5}(a) summarizes the variation in threshold currents as a function of out-of-plane field strengths at three distinct damping levels. The plots of $I_{th}$ versus $\mu_{0}H$ exhibit a parabolic-like non-monotonic trend for all three damping values. In general, the threshold current ($I_{th}$) is expected to increase monotonically with the applied magnetic field strength~\cite{Dvornik2018PRA}. However, in our case, an intriguing deviation is observed, wherein a reduction in $I_{th}$ occurs at intermediate field strengths. This reduction reaches a minimum value before the threshold current begins to increase at strong out-of-plane fields. A similar parabolic-like behavior has been reported by Yin et al.~\cite{yin2018damping} in Py${_\text{100-x-y}}$Pt${_\text{x}}$Ag$_{\text{y}}$/Pt-based nanoconstriction SHNOs. This intriguing non-monotonic behavior observed in the field dependence of threshold current underlines a more in-depth investigation of the dynamics of auto-oscillating modes in the constriction region. 

To elucidate this behavior, we performed simulations of current sweep auto-oscillations at three different operating regimes depicted in Fig. \ref{fig:5}(a). Figures \ref{fig:5}(b-d) reveal three distinct characteristics of auto-oscillation frequency vs. current, highlighting the distinct evolution of auto-oscillating modes. In weak fields, such as $\mu_{0}H$ = 0.1 T, the auto-oscillation frequency exhibits negligible change with current, and the mode remains quasi-linear, as illustrated in Fig. \ref{fig:5}(b). The spatially simulated profile, depicted as the inset of Fig. \ref{fig:5}(b), confirms that the mode originates at the constriction edges due to inhomogeneous current distributions, as shown in Fig. \ref{fig:1}(b). A small negative non-linearity  (refer to Fig. \ref{fig:2}) governs this quasi-linear mode at weak out-of-plane fields~\cite{slavin2009,Dvornik2018PRA,Fulara2019SciAdv}. As the applied field strength increases, the non-linearity becomes more negative (see Fig. \ref{fig:2}), leading to a stronger localization of the auto-oscillating mode at the constriction edges~\cite{slavin2009,gerhart2007prb}. Consequently, the mode undergoes a transition from a quasi-linear behavior to a strongly localized spin-wave bullet behavior, resulting in an abrupt drop in frequency, as shown in Figure \ref{fig:5}(c). With a further increase in the field strength, the non-linearity initially increases from a negative value at low currents, passes through a zero value, and eventually transitions to a positive value at higher currents~\cite{Dvornik2018PRA,RajabaliPRA2023} (refer to Fig. \ref{fig:2}). This behavior arises due to the magnetization vector undergoing precession at an increasingly larger angle as the current magnitude rises. Consequently, there is an effective reduction of $M_{\text{S}}$ through the projection of magnetization ($M_{\beta}$ = $M_{\text{S}}$ cos $\beta$) along the applied field direction~\cite{dvornik2018anomalous,Fulara2019SciAdv}, resulting in a shift of the non-linearity from a negative value to a positive one, as illustrated by black arrow in Fig. \ref{fig:2}. This transition occurs as the auto-oscillating mode detaches from the constriction edges and transforms into a bulk type. As shown in the inset of Fig. \ref{fig:5}(d), the auto-oscillating mode shifts inwards into the interior of the constriction and expands away from the constriction as the current increases. 

To look forward, controllable frequency tunability in SHNOs can enhance their adaptability and versatility, making them suitable for a wide range of practical applications in communication, adaptive signal processing, information storage, and computing~\cite{bonetti2009apl,romera2018nature,fulara2020natcomm,Zahedinejad2022natmat}. Depending on the desired functionality, the SHNOs can be tuned to operate in different frequency ranges (refer to Fig. \ref{fig:3}), enabling their use in a variety of applications within a single device. In the realm of neuromorphic computing, the ability to adjust the oscillation frequency of SHNOs becomes particularly valuable. This tunability can be leveraged to emulate synaptic plasticity, where frequency modulation represents variations in synaptic strength~\cite{romera2018nature,markovic2020physics}. This, in turn, facilitates the implementation of learning and memory functions within neuromorphic circuits. Moreover, the tunable frequencies of SHNOs enable them to replicate the dynamic behavior of neurons in response to changing input conditions. The capability to dynamically adjust frequencies on the fly allows for swift modifications to neural connections, facilitating continuous learning in neuromorphic systems. With controllable frequency tunability, SHNOs exhibit the capability to dynamically adapt their frequencies based on the computational task at hand. This not only optimizes energy consumption but also contributes to the overall efficiency of the system. 

\section{Conclusion}
In summary, we have studied the individual influence of key magnetodynamical parameters, namely effective magnetization ($\mu_0 M_{\text{eff}}$) and magnetic damping ($\alpha$), on the auto-oscillation characteristics of nano-constriction-based SHNOs. Our findings reveal a significant variation in the impact of effective magnetization on frequency tunability, particularly in response to applied field strengths. Conversely, magnetic damping primarily influences the threshold current, exerting minimal effects on frequency tunability. We observed a linear increase in the threshold current with increasing magnetic damping under a constant magnetic field. However, when exploring the interplay of threshold current with applied field strengths, a distinct parabolic-like trend emerges. This behavior suggests qualitatively different origins of auto-oscillating modes across different field landscapes, providing valuable insights into the intricate magnetodynamics of SHNOs in nano-constriction geometry. Our study not only contributes to oscillator-based neuromorphic computing through controllable frequency tunability but also provides valuable insights into the intricate auto-oscillation dynamics of nano-constriction-based SHNOs.
 
 \subsection*{Acknowledgements}
 
This work is supported by the Faculty Initiation Grant (FIG) sponsored by SRIC, IIT Roorkee. We thankfully acknowledge the Institute Computer Center (ICC), IIT Roorkee for providing a high-end computational facility to run simulations.
 

%

\end{document}